\documentclass[12pt,aps,prd,preprint,preprintnumbers,tightenlines,
                showpacs,nofootinbib]{revtex4}
\newcommand{\PRE}[1]{{#1}} % Use if preprint style
\usepackage{bm} \usepackage{epsfig}

\newcommand{\kev}{\text{keV}}
\newcommand{\mev}{\text{MeV}}
\newcommand{\gev}{\text{GeV}}
\newcommand{\tev}{\text{TeV}}

\newcommand{\mm}{\text{mm}}
\newcommand{\cm}{\text{cm}}
\newcommand{\m}{\text{m}}

\newcommand{\s}{\text{s}}

\newcommand{\sr}{\text{sr}}

\newcommand{\eqref}[1]{Eq.~(\ref{#1})}

\newcommand{\secref}[1]{Sec.~\ref{sec:#1}}

\newcommand{\figref}[1]{Fig.~\ref{fig:#1}}
\newcommand{\figsref}[2]{Figs.~\ref{fig:#1} and \ref{fig:#2}}

\newcommand{\kkh}{H^{\pm\, 1}}
\newcommand{\mG}{m_{G^1}}
\newcommand{\mB}{m_{B^1}}

\newcommand{\mH}{m_{\kkh}}
\newcommand{\mA}{m_{A^{1}}}
\newcommand{\trh}{T_{\text{RH}}}

\begin{document}

\preprint{UCI-TR-2006-21}

\title{
\PRE{\vspace*{1.5in}}
Exotic Collider Signals from the Complete Phase Diagram
of Minimal Universal Extra Dimensions \\
\PRE{\vspace*{0.3in}} }

\author{Jose~A.~R.~Cembranos}
\affiliation{Department of Physics and Astronomy, University of
California, Irvine, CA 92697, USA \PRE{\vspace*{.5in}} }
\author{Jonathan L.~Feng}
\affiliation{Department of Physics and Astronomy, University of
California, Irvine, CA 92697, USA \PRE{\vspace*{.5in}} }
\author{Louis E.~Strigari%
\PRE{\vspace*{.2in}} } \affiliation{Department of Physics and
Astronomy, University of California, Irvine, CA 92697, USA
\PRE{\vspace*{.5in}} }
%\date{November 2006}

\begin{abstract}
\PRE{\vspace*{.3in}} Minimal universal extra dimensions (mUED) is
often thought to predict that the lightest Kaluza-Klein particle (LKP)
is the Kaluza-Klein gauge boson $B^1$, leading to conventional missing
energy signals at colliders and WIMP dark matter.  In fact, the
implications of mUED are far richer: the $B^1$, charged Higgs boson
$H^{\pm\, 1}$, and graviton $G^1$ are all possible LKPs, leading to
many different ``phases'' with distinct signatures.  Considering the
complete phase diagram, we find predictions for charged or neutral
particles with decay lengths of microns to tens of meters; WIMP,
superWIMP, or charged relic particles; metastable particles with
lifetimes of the order of or in excess of the age of the universe; and
scenarios combining two or more of these phenomena.  In the
cosmologically preferred region, the Higgs boson mass is between 180
and 245 GeV, the LKP mass is between 810 and 1400 GeV, and the maximal
splitting between first Kaluza-Klein modes is less than 320 GeV.  This
region predicts a variety of exotic collider signals, such as slow
charged particles, displaced vertices, tracks with non-vanishing
impact parameters, track kinks, and even vanishing charged tracks, all
of which provide early discovery possibilities at the Large Hadron
Collider.
\end{abstract}

\pacs{11.10.Kk, 12.60.-i, 95.35.+d, 98.80.Cq}
%11.10.Kk   Field theories in dimensions other than four
%12.60.-i   Models beyond the standard model
%95.35.+d   Dark matter
%98.80.Cq   Particle-theory and field-theory models of the early 
%              Universe (including cosmic pancakes, cosmic strings, 
%              chaotic phenomena, inflationary universe, etc.) 

\maketitle

\section{Introduction}
\label{sec:introduction}

The idea that there may be extra spatial dimensions is an old one,
going back at least as far as the work of Kaluza and Klein in the
1920's~\cite{KK}.  Their original idea was untenable, but it has many
modern descendants, of which the closest living relative is universal
extra dimensions (UED)~\cite{Appelquist:2000nn}.  In UED, all
particles propagate in flat, compact extra dimensions of size
$10^{-18}~\m$ or smaller.  Each known particle has an associated set
of heavy partner particles, providing a wealth of possible
implications for particle physics and cosmology.

In this study we consider minimal UED (mUED) in which there is one
extra dimension of size $R$ compactified on an $S^1/Z_2$ orbifold,
where $Z_2$ is the action $y \to -y$, with $y$ the coordinate of the
extra dimension.  Every state of the standard model has a partner
particle at Kaluza-Klein (KK) level $n$ with mass $nR^{-1}$,
supplemented by tree-level contributions from electroweak symmetry
breaking and radiative corrections~\cite{Cheng:2002iz,Cheng:2002ab}.
In general UED theories, there may also be contributions to the KK
masses from mass terms localized on the orbifold boundaries.  These
contributions would generically violate bounds on flavor and CP
violation.  To remain consistent with experiment, a simple assumption,
which defines mUED, is that these boundary contributions are absent.
The resulting model preserves a discrete parity known as KK-parity,
which implies that the lightest KK particle (LKP) is stable and a
possible dark matter
candidate~\cite{Servant:2002aq,Cheng:2002ej,Feng:2003xh,%
Feng:2003uy,Feng:2003nr}.

Minimal UED is therefore an extremely simple, viable extra dimensional
extension of the standard model.  It is completely determined by only
2 parameters: $m_h$, the mass of the standard model Higgs boson, and
one new parameter, $R$, the compactification radius.  (In detail,
there is also a third parameter, the cutoff scale $\Lambda$, but the
dependence on $\Lambda$ is logarithmic and weak, as discussed below.)
Precision electroweak measurements require $R^{-1} \agt
250~\gev$~\cite{Appelquist:2000nn,Appelquist:2002wb}, with other low
energy constraints similar or
weaker~\cite{Agashe:2001xt,Appelquist:2001jz}.  Particle physics alone
does not place an upper bound on $R^{-1}$, but the thermal relic
density of LKPs grows with $R^{-1}$, and LKPs would overclose the
universe for $R^{-1} > 1.5~\tev$~\cite{Servant:2002aq,%
Kakizaki:2005en,Kakizaki:2005uy,Burnell:2005hm,Kong:2005hn,%
Kakizaki:2006dz}, providing strong motivation for considering
weak-scale KK particles.  For the Higgs boson mass, the direct
constraints on the standard model also apply to UED and require $m_h >
114.4~\gev$ at 95\% CL~\cite{Barate:2003sz}.  In contrast, however,
the indirect bounds on $m_h$ are significantly weakened relative to
the standard model, requiring only $m_h < 900~\gev$ for $R^{-1} =
250~\gev$ and $m_h < 300~\gev$ for $R^{-1} = 1~\tev$ at 90\%
CL~\cite{Appelquist:2002wb}.

Early studies of UED focused on the line in model parameter space
defined by $m_h = 120~\gev$~\cite{Cheng:2002ab} and neglected the
existence of the KK graviton $G^1$~\cite{Servant:2002aq,Cheng:2002ej}.
Given these assumptions, for $R^{-1} \agt 250~\gev$, the LKP is the
hypercharge gauge boson $B^1$, and these studies therefore focused on
missing energy signals at colliders and weakly-interacting massive
particle (WIMP) dark matter for cosmology.  These predictions are
similar to those from supersymmetry with $R$-parity conservation. UED
with KK-parity and supersymmetry with $R$-parity predict different
collider event rates for similar spectra, and the different spins of
partner particles may be distinguished through, for example, indirect
dark matter detection in positrons~\cite{Cheng:2002ej}.  Nevertheless,
the difficulty of distinguishing UED and supersymmetry has attracted
much attention and been a fertile testing ground for future
experiments, especially the Large Hadron Collider
(LHC)~\cite{Macesanu:2005jx}.

In fact, however, more recent studies have shown that framework of UED
is far richer than indicated above.  First, it was noted that the KK
graviton $G^1$ necessarily exists in any UED model and may be the LKP,
leading not to WIMP dark matter, but to superWIMP dark matter, with a
completely different set of cosmological and astroparticle
signatures~\cite{Feng:2003xh,Feng:2003uy,Feng:2003nr}.  Second,
studies have now emphasized that, by relaxing the constraint $m_h =
120~\gev$ and considering higher values, KK Higgs bosons may become
lighter than the $B^1$.  That both of these possibilities may be
realized in a general UED model is, perhaps, not surprising.
Remarkably, however, all of these complexities arise even in the
extremely constrained framework of mUED.  Any one of the $G^1$, $B^1$,
and the charged Higgs boson $\kkh$ may be the LKP, leading to many
different ``phases'' of parameter space with qualitatively distinct
signatures.  The ``triple point,'' where $m_{G^1} = m_{B^1} =
m_{\kkh}$, lies in the heart of parameter space at $(R^{-1}, m_h)
\approx (810~\gev, 245~\gev)$, leading to many interesting features.

With this as motivation, we consider here the full parameter space of
mUED and its implications for particle physics and cosmology.  In
\secref{masses} we present the complete phase diagram of mUED. For
reasons given below, we begin by excluding the graviton $G^1$ from
consideration and define phases to be regions with distinct standard
model (NLKP, LKP) pairs.  With this classification, we explore the
collider physics of mUED in \secref{lifetimes}, and find that
long-lived particles with macroscopic decay lengths at colliders are
common in the full parameter space.

In \secref{cosmology}, we then include the graviton $G^1$, and examine
each phase in light of cosmological constraints on charged dark
matter, diffuse photon spectra, and dark matter thermal relic
densities.  We find that each phase of parameter space is
cosmologically viable, given, for example, a low enough reheat
temperature $\trh > 1~\mev$, justifying the effort made in
\secref{lifetimes} to elucidate the collider implications of every
phase.  At the same time, for standard cosmological scenarios with
$\trh \agt 10~\gev$, we find that the viable region of parameter space
has $180~\gev \alt m_h \alt 245~\gev$, $810~\gev \alt R^{-1} \alt
1400~\gev$, and a maximal splitting between the LKP and the heaviest
$n=1$ KK state (the KK gluon $g^1$) always less than 320 GeV.  In
addition, this cosmologically favored region of the phase diagram
predicts charged particle decays, such as $H^{+\, 1} \to B^1 u
\bar{d}, B^1 c \bar{s}, B^1 e^+ \nu_e, B^1 \mu^+ \nu_{\mu}, B^1 \tau^+
\nu_{\tau}$, with macroscopic decay lengths from microns to tens of
meters, leading to the possibility of spectacular signals and early
discoveries at the LHC. These predictions are rather striking and
differentiate mUED from supersymmetry and essentially all other
frameworks for new physics proposed to date.  These collider signals,
as well as other conclusions and future directions, are presented in
\secref{summary}.

Finally, our conventions and notations are collected in
Appendix~\ref{sec:appendix}, along with Feynman rules and other
technical details helpful for determining decay widths.

\section{Mass Spectrum and Phase Diagram}
\label{sec:masses}

As we will see below, although mUED is among the simplest extra
dimensional extensions of the standard model, the spectrum of mUED is
remarkably intricate.  We will find that there are several LKP
candidates, and degeneracies $\alt 1~\gev$ among the lightest KK
states are not uncommon.

Rather than deal immediately with the complexity of this complicated
spectrum, we begin in this section by ignoring the existence of the KK
graviton $G^1$.  This is beneficial for two reasons.  First, because
the KK graviton has only cosmological significance, neglecting it
allows us to defer cosmological considerations and the accompanying
dependence on early universe assumptions to focus on collider physics
predictions, which are much more robust. Second, this simplification
allows us to divide the parameter space into a manageable number of
phases with qualitatively distinct signatures at colliders.

The definition of ``phase'' is, of course, somewhat arbitrary.  The
simplest option is to divide the parameter space into regions with
different LKPs, as a pair of LKPs is produced in every KK event, and
so the nature of the LKP determines to a large extent the collider
signatures.  At the other extreme, one might argue that, viewed in
sufficient detail, collider signals depend on the entire KK spectrum,
making each point in parameter space a different phase and eliminating
the utility of the concept of phases in model parameter space.

For mUED, however, we find that a useful definition of phases lies
between these two extremes.  As we will see, in mUED, the nature of
both the LKP and the next-to-lightest KK particle (NLKP) are
important, as they both impact {\em qualitatively} what signatures are
predicted.  In this section, we therefore, exclude the $G^1$ from
consideration and divide the parameter space into phases with distinct
standard model (NLKP, LKP) pairs.  In mUED, and without taking into
account the KK graviton, the following KK particles may be either the
LKP or the NLKP: the hypercharge gauge boson $B^1$, the 3 SU(2)
singlet leptons $e_R^1$, $\mu_R^1$, and $\tau_R^1$, the charged Higgs
boson $H^{\pm\, 1}$, and the CP-odd Higgs boson $A^1$.  The complete
spectrum for mUED, including one-loop corrections, was first presented
in Ref.~\cite{Cheng:2002iz}.  Here we reproduce the formulae that
determine the masses of these states.

The mUED spectrum is completely determined by 3 parameters,
\begin{equation}
R^{-1} , \ m_h , \ \Lambda \ ,
\end{equation}
where $R$ is the compactification radius, $m_h$ is the Higgs boson
mass, and $\Lambda$ is the cutoff scale.  As seen below, masses depend
only logarithmically on $\Lambda$.  The dependence is therefore weak,
and we have checked that, for the range $10 \le \Lambda R \le 50$, our
main results are essentially independent of $\Lambda$.  For numerical
results, we take $\Lambda R = 20$ throughout this study.

The $G^1$, $\kkh$, and $A^1$ masses are
\begin{eqnarray}
\mG &=& R^{-1} \\
\mH^2 &=& R^{-2} + m_W^2 + \delta m_H^2 \\
\mA^2 &=& R^{-2} + m_Z^2 + \delta m_H^2 \ ,
\end{eqnarray}
where the radiative correction to the Higgs boson masses is
\begin{eqnarray}
\delta m_H^2 &=& \left( \frac{3}{2} g^2 + \frac{3}{4} g'^2 -
\frac{m_h^2}{v^2} \right) \frac{\ln(\Lambda^2 R^2)}{16
    \pi^2} \, R^{-2} \ ,
\end{eqnarray}
and $v \simeq 246~\gev$ is the Higgs boson vacuum expectation value.
Note that the electroweak symmetry breaking and radiative corrections
to $\mG$ are negligible, and the corrections to $\mH$ and $m_{A^1}$
are not only small, but may also be either positive or negative,
depending on $m_h$.

The KK charged lepton mass matrix, where $l_L$ and $l_R$ denote SU(2)
doublet and singlet states, respectively, is
\begin{equation}
\left( \begin{array}{cc}
R^{-1} + \delta m_{l_L^1} & m_l \\
m_l & - R^{-1} - \delta m_{l_R^1} 
\end{array} \right) \ ,
\label{lmatrix}
\end{equation}
where 
\begin{eqnarray}
\delta m_{l_L^1} &=& \left( \frac{27}{16} g^2 + \frac{9}{16} g'^2 \right)
\frac{\ln ( \Lambda^2 R^2)}{16 \pi^2} \, R^{-1} \\
\delta m_{l_R^1} &=& \frac{9}{4} g'^2 
\frac{\ln ( \Lambda^2 R^2)}{16 \pi^2} \, R^{-1} \ .
\end{eqnarray}
For the viable regions of parameter space, the lighter eigenstate is
very nearly a pure $l_R^1$ state, and so we refer to it as $l_R^1$.
We diagonalize the mass matrix of \eqref{lmatrix} (and also the matrix
of \eqref{BWmatrix} below) in obtaining numerical results, but the
$l_R^1$ mass is very well-approximated by the lower right-hand entry
of the mass matrix.  The KK leptons $e_R^1$, $\mu_R^1$, $\tau_R^1$ are
extremely degenerate, with $m_{\tau_R^1\, (\mu_R^1)} - m_{e_R^1}
\approx m_{\tau\, (\mu)}^2 R \sim 1~\mev\, (10~\kev)$.

The neutral electroweak gauge boson masses are, in the basis $(B^1,
W^1)$, 
\begin{equation}
\left( 
\begin{array}{cc}
R^{-2} + \frac{1}{4} g'^2 v^2 + \delta \mB^2 &
\frac{1}{4} g' g v^2 \\
\frac{1}{4} g' g v^2 &
R^{-2} + \frac{1}{4} g^2 v^2 + \delta m_{W^1}^2 
\end{array}
\right) \ ,
\label{BWmatrix}
\end{equation}
where 
\begin{eqnarray}
\delta \mB^2 &=& \left( - \frac{39}{2} \frac{g'^2 \zeta(3)}{16 \pi^4}
- \frac{g'^2}{6} \frac{\ln ( \Lambda^2 R^2)}{16 \pi^2} \right) R^{-2} \\
\delta m_{W^1}^2 &=& \left( - \frac{5}{2} \frac{g^2 \zeta(3)}{16 \pi^4}
+ \frac{15g^2}{2} \frac{\ln ( \Lambda^2 R^2)}{16 \pi^2} \right) R^{-2} 
\ ,
\end{eqnarray}
and $\zeta$ is the Riemann zeta function, with $\zeta(3) \simeq
1.202$.  For the viable regions of parameter space, the lighter
eigenstate is approximately a pure $B^1$ state, and so we refer to it
as $B^1$.  Its mass is well approximated by the upper left-hand entry
of the mass matrix.  Note that the contribution from electroweak
symmetry breaking may be canceled by the radiative correction.

\begin{figure}
\resizebox{3.56in}{!}{
\includegraphics{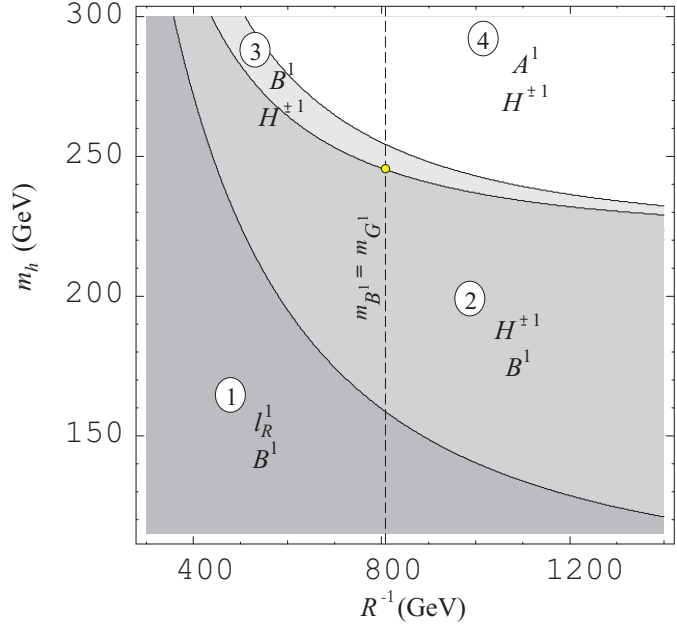}
} 
\caption{The complete collider phase diagram of mUED in the $(R^{-1},
m_h)$ plane, where $R$ is the compactification radius, and $m_h$ is
the Higgs boson mass.  The KK graviton $G^1$ has been excluded from
consideration, and the standard model (NLKP, LKP) pairs in each phase
are as indicated.  We have set $\Lambda R = 20$.  For reference, the
line on which $m_{G^1} = m_{B^1}$ is also plotted.  The ``triple
point,'' where $m_{B^1} = m_{G^1} = \mH$, is at $(R^{-1}, m_h) \approx
(810~\gev, 245~\gev)$.
\label{fig:phases} }
\end{figure}

For the reasons given above, we now ignore the $G^1$ and divide the
parameter space into regions with distinct (NLKP, LKP) pairs.  The
result is given in \figref{phases}.  There are four phases, which,
from lower left to upper right, have the (NLKP, LKP) combinations
1:$(l_R^1, B^1)$, 2:$(\kkh, B^1)$, 3:$(B^1, \kkh)$, and 4:$(A^1,
\kkh)$.  Note that the line $m_h = 120~\gev$ lies completely in Phase
1 (for $115~\gev \alt R^{-1} \alt 1430~\gev$), but for larger $m_h$,
the $\kkh$ becomes the NLKP.  For even larger $m_h$, the $\kkh$
becomes the LKP, and for still larger $m_h$, the lightest two KK
particles are the two Higgs bosons $A^1$ and $\kkh$.  Although we are
temporarily excluding the KK graviton from consideration here, for
later reference, we have also plotted the line on which $m_{G^1} =
m_{B^1}$.  The ``triple point,'' where $m_{B^1} = m_{G^1} = \mH$, is
at $(R^{-1}, m_h) \approx (810~\gev, 245~\gev)$ for $\Lambda R = 20$,
and its location is essentially independent of $\Lambda$ in the range
$10 \le \Lambda R \le 50$.

\begin{figure}
\resizebox{3.56in}{!}{
\includegraphics{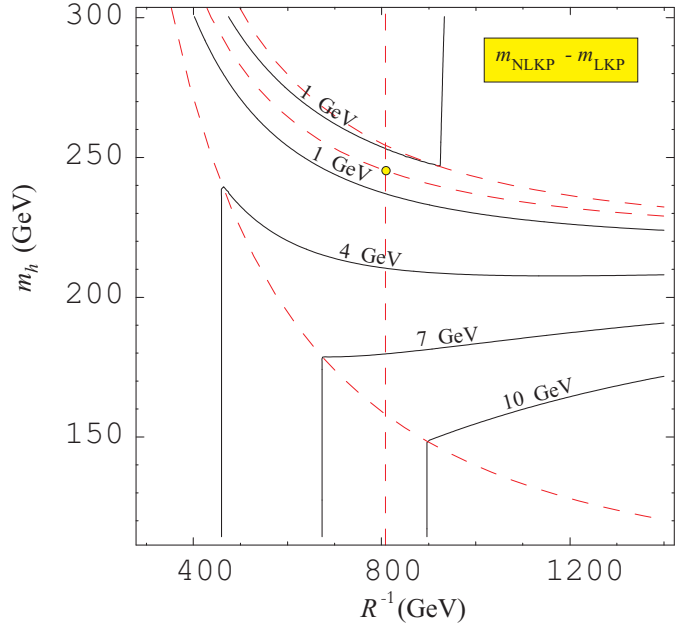}
} 
\caption{Mass splittings (in GeV) between the standard model NLKP and
LKP in the full phase diagram of mUED. The dashed lines are the
boundaries between different phases shown in \figref{phases}.  We have
set $\Lambda R = 20$.
\label{fig:splittings} }
\end{figure}

In \figref{splittings}, we show the mass splittings $\Delta m =
m_{\text{NLKP}} - m_{\text{LKP}}$ in the full parameter space.
Remarkably, the mass splittings are only of the order of 1 to 10 GeV
throughout the full phase diagram.  One might expect splittings of the
order of $R^{-1} / (16 \pi^2), m_W^2 R \sim 10~\gev$; modest
additional cancellations arising from effects highlighted above in fact
make this an overestimate in most of parameter space.

\section{Long-lived particles at Colliders}
\label{sec:lifetimes}

\begin{figure}
\resizebox{4.56in}{!}{
\includegraphics{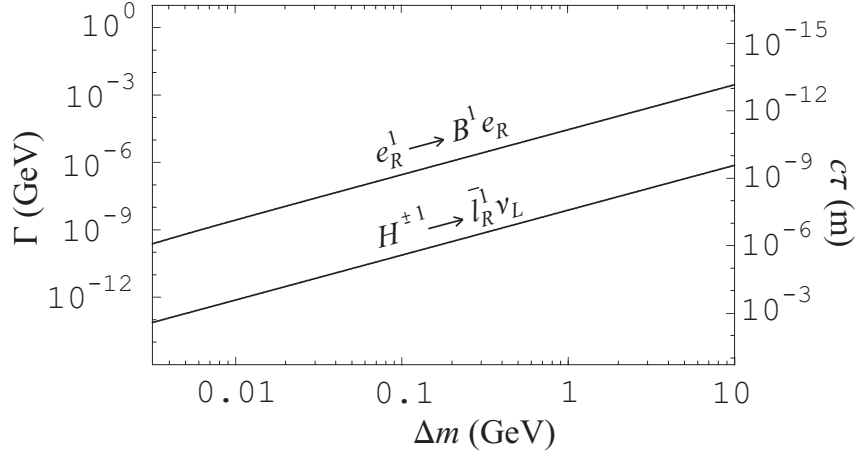}
} 
\caption{Decay widths and decay lengths as a function of the mass
  splitting $\Delta m$ between KK states for the two-body decays
  indicated. We have fixed the decaying particle's mass to $M =
  1~\tev$. 
\label{fig:twobody} }
\end{figure}

\begin{figure}
\resizebox{3.56in}{!}{
\includegraphics{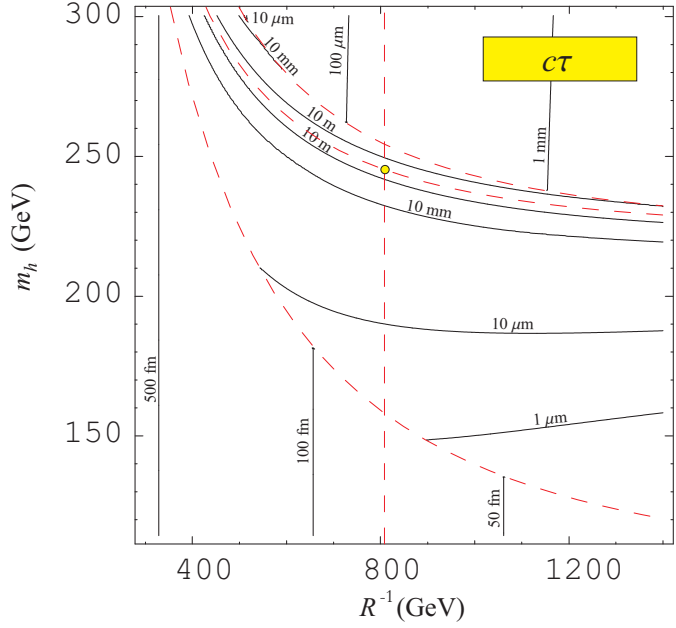}
} 
\caption{Standard model NLKP decay lengths in the full phase space of
mUED. We have set $\Lambda R = 20$ and neglected the KK graviton
$G^1$. The dashed curves are the boundaries of phases shown in
\figref{phases}.
\label{fig:lifetimes} }
\end{figure}

In \secref{masses}, we found fractional degeneracies of 0.001 to 0.01
throughout the mUED phase diagram.  These degeneracies suppress NLKP
decay widths, such that NLKPs produced in colliders may decay at
points macroscopically separated from the interaction point.  In this
section we present numerical results for the most relevant decays.
Two- and three-body decay widths and lengths are given in
\figsref{twobody}{threebody}, respectively, and the NLKP decay lengths
throughout mUED parameter space are given in \figref{lifetimes}.
Analytical formulae for the decay widths, along with useful results
for calculating them, are given in Appendix~\ref{sec:appendix}.

In Phase 1, the NLKP to LKP decay is $l^1_R \to B^1 l_R$, which is a
two-body decay suppressed only by the mass degeneracy discussed above.
As can be seen in \eqref{decaylb} and \figref{twobody}, for mass
splittings $100~\mev \alt \Delta m \alt 10~\gev$ the decay lengths are
$10^{-9}~\m \agt c\tau \agt 10^{-13}~\m$ (for $R^{-1} = 1~\tev$).  To
observe displaced vertices or non-vanishing impact parameters at
colliders, decay lengths should be greater than about $10~\mu\m$.  The
$l^1_R \to B^1 l_R$ decay lengths are, then, too short to be
observable in any part of Phase 1.

\begin{figure}
\resizebox{4.56in}{!}{
\includegraphics{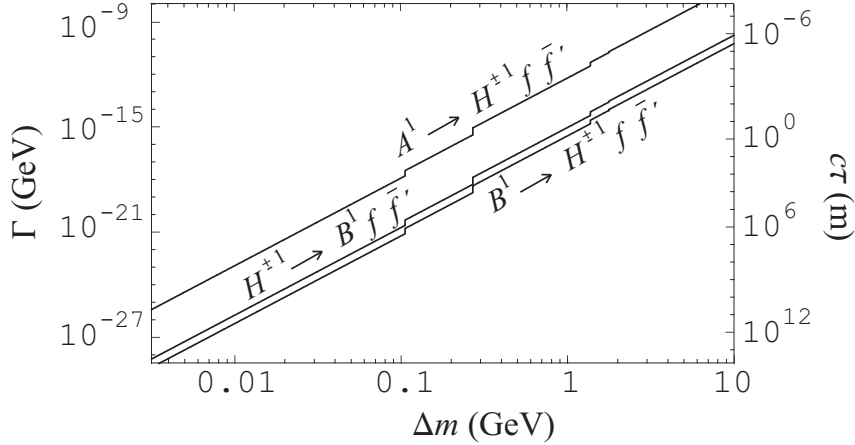}
} 
\caption{Decay widths and decay lengths as a function of the mass
  splitting $\Delta m$ between KK states for the three-body decays
  indicated.  We have fixed the decaying particle's mass to $M =
  1~\tev$. The discontinuities result from setting $m_f = m_{\bar{f}'}
  = 0$ for all kinematically accessible final states.  We include the
  $ud$ final state above the $\pi \pi$ threshold.
\label{fig:threebody} }
\end{figure}

Are there other particles in Phase 1 that can have macroscopic decay
lengths?  Although the answer is no, this question merits discussion.
The 4 lightest standard model states in most of Phase 1 are, in
decreasing order of mass, $(A^1, \kkh, l_R^1, B^1)$. The charged Higgs
boson decays through $\kkh \to B^1 f \bar{f}'$ and $H^{+\, 1} \to
\bar{l}_R^1 \nu_l$.  The former is three-body and parametrically
suppressed by $(\Delta m)^5 / (m_W^2 M^2)$.  As can be seen in
\eqref{decayhb} and \figref{threebody}, this by itself would lead to a
macroscopic decay length.  The latter decay is two-body, but
suppressed by Yukawa couplings.  Unfortunately, this suppression for
the $\tau$ decay mode is insufficient to keep the decay length
macroscopic.  Of course, there is an extremely thin region in Phase 1
along the border between Phase 1 and 2 in which $H^{+\, 1} \to
\bar{\mu}_R^1 \nu_{\mu}$ is kinematically allowed but $H^{+\, 1} \to
\bar{\tau}_R^1 \nu_{\tau}$ isn't, which would make the $\kkh$ decay
length observably long, but we do not consider this further.

In Phase 1 the $A^1$ may decay through $A^1 \to \kkh f \bar{f}' , B^1
\gamma, \bar{l}_R^1 l_R$.  (Note that $A^1 \to B^1 Z^* \to B^1 f
\bar{f}$ is not allowed, because the tree-level $AZZ$ coupling is
forbidden by CP-invariance, and, of course, the tree-level $AZ\gamma$
coupling is absent because the $A$ is neutral.)  The first two are
highly suppressed; $A^1 \to \kkh f \bar{f}'$ is parametrically
suppressed by $(\Delta m)^5/m_W^4$, a huge suppression given the
degeneracies of mUED, and $A^1 \to B^1 \gamma$ is a two-body decay,
but is loop-suppressed.  However, $\Gamma(A^1 \to \bar{l}_R^1 l_R) =
\Gamma(\kkh \to \bar{l}_R^1 \nu_l)$, and so, once again, the Yukawa
coupling decay is not sufficiently suppressed to produce a long-lived
track.

We conclude, then, that there are no long-lived tracks in Phase 1.  It
is rather remarkable, however, that the NNLKP and the NNNLKP can be so
close to being long-lived, despite the many decay channels open to
them.

In Phase 2 the NLKP to LKP decay, $\kkh \to B^1 f \bar{f}'$, is
three-body.  The parametric phase space suppression of $(\Delta m)^5 /
(m_W^2 M^2)$ leads to decay lengths of 20 cm for $\Delta m = 1~\gev$,
as seen in \eqref{decayhb} and \figref{threebody}.  As a result,
displaced vertices and non-zero impact parameters are expected for
much of Phase 2 from the decays $H^{+\, 1} \to B^1 e^+_R \nu_e, B^1
\mu^+_R \nu_{\mu}, B^1 \tau^+_R \nu_{\tau}, B^1 u \bar{d}, B^1 c
\bar{s}$.  The standard model fermions produced are extremely soft,
creating a difficult challenge for collider experiments.  We discuss
these issues in \secref{summary}.

Approaching the upper boundary of Phase 2, the $H^{+\, 1}$ and $B^1$
may be arbitrarily degenerate.  For $\Delta m \alt 0.4~\gev$, the
$H^{+\, 1}$ is essentially stable for collider phenomenology (again,
see \eqref{decayhb} and \figref{threebody}), resulting in signals
associated with slow, metastable charged particles, such as
highly-ionizing tracks and time-of-flight signals.

In Phase 3, the $B^1$ and $\kkh$ exchange roles relative to Phase 2,
and the NLKP decay is $B^1 \to \kkh f \bar{f}'$.  For small mass
splittings $\Delta m \ll M$, this decay width differs from that for
$\kkh \to B^1 f \bar{f}'$ only by the average over 3 initial spin
polarizations, and so the decay length is 3 times longer, as shown in
\eqref{decaybh} and \figref{threebody}.  This decay length is longer
than 10 mm for most of Phase 3, and so the $B^1$ is potentially
observable as a long-lived particle.  Note, however, that in contrast
to Phase 2, the parent particle here is neutral and the heavy daughter
particle is charged. 

As one approaches the Phase 3/Phase 2 boundary in Phase 3, of course,
the splitting between the $B^1$ and $H^{+\, 1}$ may be arbitrarily
small. For $\Delta m \alt 0.5~\gev$, the $B^1$ lifetime is so large
that one expects the standard missing energy signal.
 
Finally, in Phase 4 the NLKP decay is $A^1 \to \kkh f \bar{f}'$; the
decay length is given in \eqref{decayah} and \figref{threebody}.  The
$A^1$ and $\kkh$ masses are similarly controlled by $R^{-1}$ and
$m_h$, and so their mass splitting is almost constant throughout the
phase diagram.  In Phase 4, $\Delta m = m_{A^1} - m_{\kkh}$ varies
from 0.7 GeV to 1.8 GeV, and the $A^1$ decay length varies from
$10~\mu\m$ to 1 mm.  This prediction is quite robust, and the decay
length is therefore likely in the observable range.

\section{Cosmological Constraints}
\label{sec:cosmology}

So far we have neglected the KK graviton $G^1$ and ignored
cosmological constraints.  In this section, we re-introduce $G^1$ and
discuss the three most stringent cosmological bounds on mUED: null
searches for exotic charged stable particles, the diffuse photon
spectrum, and the thermal relic density of WIMP dark matter.

Throughout this section, we assume that $G^1$ production during
reheating after inflation is negligible.  This may very well {\em not}
be the case, as $G^1$ production is extremely effective in
UED~\cite{Feng:2003nr}, and may be significant even for reheating
temperatures $\trh \sim \tev$.  The possibility of KK graviton
production during reheating has been examined in
Refs.~\cite{Feng:2003nr,Shah:2006gs}.

\begin{figure}
\resizebox{3.56in}{!}{
\includegraphics{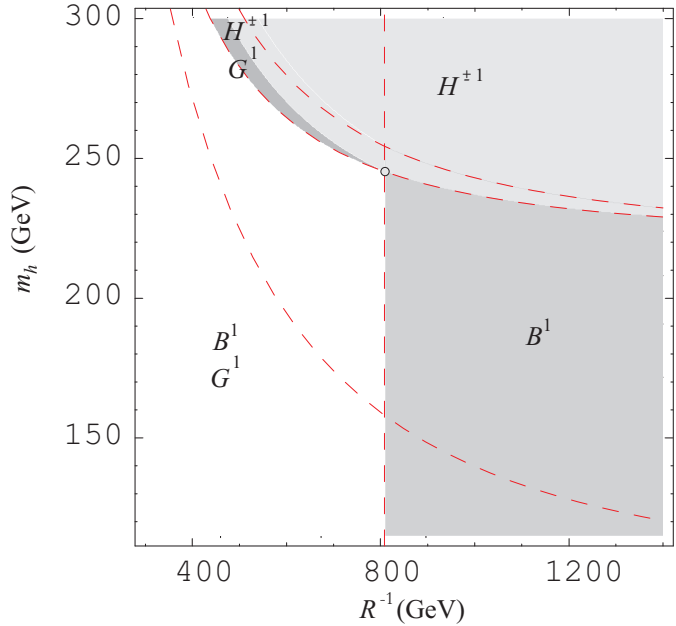}
} 
\caption{The complete cosmological phase diagram of mUED in the
$(R^{-1}, m_h)$ plane.  Phases are determined by the LKP when it is a
standard model KK particle ($H^{+\, 1}$ or $B^{1}$), and by the (NLKP,
LKP) combination when the LKP is the KK graviton $G^1$.  The dashed
lines denote boundaries of the collider phase diagram, as given in
\figref{phases}.  We have set $\Lambda R = 20$. 
\label{fig:graviton} }
\end{figure}

With this assumption, the KK graviton is cosmologically relevant when
it is the LKP.  In \figref{graviton}, we present the complete
cosmological phase diagram of mUED, where phases are defined by the
LKP when it is a standard model KK particle, and by the (NLKP, LKP)
combination when the LKP is the KK graviton $G^1$.

\subsection{Charged Stable Particles}
\label{sec:charged}

In Phases 3 and 4, the lightest standard model KK particle is the
charged Higgs boson $\kkh$.  Including the KK graviton $G^1$, we find
that the $\kkh$ is not the lightest KK particle everywhere in this
region: in a thin region along the lower left border of Phase 3, $G^1$
is lighter than $\kkh$ by a GeV or less (for $R^{-1}\agt 300~\gev$).
The lifetime for $\kkh \to G^1 f \bar{f}'$ is given in
\eqref{decayhg}; parametrically, it depends on $(\Delta m)^7/(m_W^2
M_4^2 \mH^2)$, where $M_4 \simeq 1.72 \times 10^{18}~\gev$ is the
4-dimensional Planck mass.  It is therefore extraordinarily suppressed
and typically many orders of magnitude greater than the age of the
Universe.  Effectively, then, the $\kkh$ is stable throughout Phases 3
and 4. (It is interesting to note, however, that this feature can
easily change if the mUED is minimally extended to include KK
right-handed neutrinos~\cite{Matsumoto:2006bf}.  In these scenarios,
the KK right-handed neutrinos effectively act as a new superWIMPs, and
can ameliorate cosmological and astrophysical constraints.)

The possibility of a stable charged particle is often considered to be
completely excluded by bounds on charged dark matter.  On the other
hand, it is also true that such particles can be diluted away to
insignificant abundances by a period of inflation.  The tension
between these two statements may be put on a quantitative footing by
asking the question in the following way: Assume inflation diluted the
$\kkh$ density to zero.  The universe then reheats, however, and Higgs
bosons $\kkh$ are regenerated.  What is the maximal reheating
temperature $\trh$ such that the resulting $\kkh$ density is
consistent with bounds on charged relic particles?  The lower the
maximal $\trh$ is, the more cosmologically disfavored the scenario.
If the maximal $\trh$ is less than 1 MeV, the required reheat
temperature is inconsistent with big bang nucleosynthesis and the
scenario may be considered excluded by cosmology.

Bounds on $\trh$ were considered in exactly this context in
Ref.~\cite{Kudo:2001ie}.  For stable charged particles $X$ with masses
$100~\gev \alt m_X \alt 1~\tev$, this study found that the extremely
stringent bound $n_X / n_H \alt 10^{-28}$~\cite{Smith:1982qu} on the
number density of charge $+1$ $X$ particles relative to that of
Hydrogen atoms in sea water requires $\trh \alt 1~\mev$, effectively
excluding such particles.  However, for $1~\tev \alt m_X \alt
1.6~\tev$, the experimental limit weakens drastically to $n_X / n_H
\alt 4 \times 10^{-17}$~\cite{Yamagata:1993jq}, and $\trh$ as high as
$\sim 1~\gev$ is possible.  Although $\trh \sim \gev$ is still
extremely low from a model building point of view, we know little
about the Universe at temperatures above 1 MeV, and such a possibility
cannot be excluded.

We conclude that Phases 3 and 4 with $R^{-1} \alt 1~\tev$ is excluded
cosmologically, but the rest of Phases 3 and 4 is allowed, provided
the reheating temperature after inflation satisfies $\trh \alt \gev$.
Of course, in all cases, the $\kkh$ relic density is insufficient to
be a significant amount of dark matter.  

\subsection{Diffuse Photon Flux}
\label{sec:diffuse}

In Phases 1 and 2 with $R^{-1} < 810~\gev$, the KK graviton $G^1$ is
lighter than the lightest standard model KK particle, the $B^1$.  The
decay $B^1 \to G^1 \gamma$ is gravitational and the $B^1-G^1$ mass
splitting is typically a few GeV or less.  The resulting decays, given
in \eqref{decaybg}, are therefore extremely suppressed. For $G^1$
masses greater than 300 GeV, the maximum mass splitting between the
$G^1$ and $B^1$ is about 1.5 GeV. Equation~(\ref{decaybg}) then
implies that the $B^1$ will decay after matter-radiation equality, and
these decays are thus strongly constrained by cosmological
observations.

For injected energies $\sim $ GeV in the redshift range of interest
($z \sim 10^3 -0$), the decay photons redshift, but the flux of
photons is otherwise unattenuated by scattering processes with the
intergalactic medium or cosmic microwave background
photons~\cite{Chen:2003gz}. The fact that the $B^1 \rightarrow G^1
\gamma$ decays fall in this transparency window is quite unique. If
the decay photons were injected with energies greater than 10 to 1000
GeV, depending on the decay redshift, a large optical depth would
result from pair-production off of cosmic microwave background
photons. For lower injected energies below 10 keV, decay photons would
lose energy from scattering off of free electrons and atoms. From
\eqref{decaybg}, a $B^1$ decaying today will produce photons with
injected energy $\sim 20~\mev$.

Constraints on late decaying $B^1$ particles have been considered
in the context of the superWIMP dark matter scenario in
Ref.~\cite{Feng:2003nr}.  The diffuse photon spectrum from $B^1 \to
G^1 \gamma$ decays is~\cite{Feng:2003nr}
\begin{equation}
\frac{d \phi}{d E} = \frac{3}{8 \pi}†
\frac{N_{\text{in}}}{V_0 E_{\text{in}}}
\ \frac{t_0}{\tau} \
\left[ \frac{E}{E_{\text{in}}} \right]^{1/2}
e^{-(E/E_{\text{in}})^{3/2}(t_0/\tau)} \Theta (E_{\text{in}} - E) \ ,
\label{diffusephoton}
\end{equation}
where $N_{\text{in}}$ is the number of $B^1$ particles at freeze-out,
$V_0$ and $t_0$ are the present volume and age of the Universe,
respectively, and $\tau$ is the $B^1$ lifetime.  $E_{\text{in}} =
(m_{B^1}^2 - m_{G^1}^2)/2m_{B^1}$ is the initial energy of the
produced photons, which is related to the present energy through
$E_{\text{in}} = (1+z_{\text{in}}) E$, where $z_{\text{in}}$ is the
redshift when produced. For $\Delta m = m_{B^1} - m_{G^1} \ll
m_{B^1}$, $\tau \propto (\Delta m)^{-3}$ and $E_{\text{in}} \approx
\Delta m$, and both are independent of the overall KK mass scale.

We compare this diffuse photon spectrum to the diffuse background
spectrum in the MeV regime from COMPTEL~\cite{Sreekumar:1997yg},
\begin{equation}
\frac{d \phi}{d E} \simeq 1.1 \times 10^{-4}
\left[ \frac{E}{5~\mev}
   \right]^{-2.4}~\mev^{-1}~\cm^{-2}~\s^{-1}~\sr^{-1} \ ,
\label{COMPTEL}
\end{equation}
valid over an energy range $E \sim 0.8-30~\mev$.  To apply the
constraints from the observations, we demand that the integrated flux
from $B^1$ decays be less than the integrated flux from the observed
background over the energy range of interest. In principle, this bound
could be made more strict by comparing the spectrum of photons from
decays to the power-law spectrum in \eqref{COMPTEL}. However, since
the photon background excludes such a large class of models, we find
the integrated flux bound an accurate criterion for our purposes.

\begin{figure}
\resizebox{3.56in}{!}{
\includegraphics{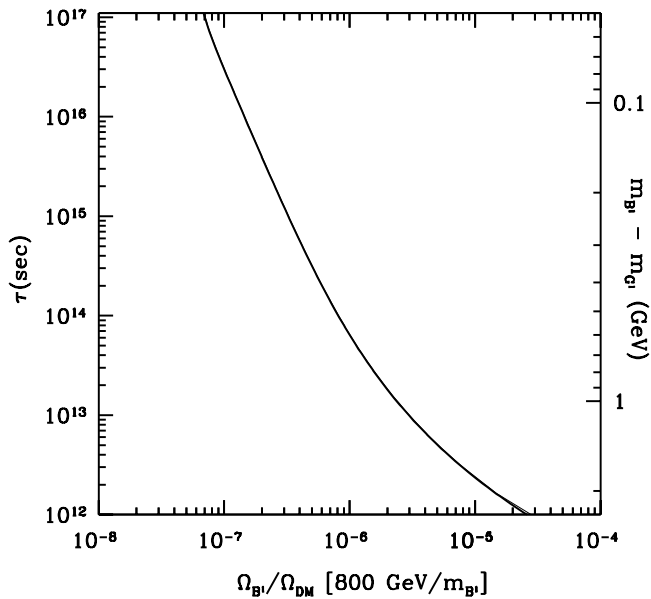}
}
\caption{Constraints on the $B^1$ relic density (if they had not
decayed) from requiring that the integrated flux of photons from $B^1
\to G^1 \gamma$ not exceed the observed MeV diffuse photon flux. The
region above the curve is excluded.
\label{fig:diffuseflux} }
\end{figure}

For a given number density $N_{\text{in}}$ the diffuse photon flux
excludes certain lifetimes $\tau$. These constraints are given in
\figref{diffuseflux}, where we express the value of $N_{\text{in}}$ in
terms of $\Omega_{B^1}$ normalized to $\Omega_{\text{DM}}$ for
$m_{B^1} = 800~\gev$.  All parameters above the curve are excluded by
our criterion.

In the relevant region of the mUED phase diagram (Phase 1 and 2 with
$R^{-1} < 810~\gev$), $\tau \agt 10^{12}$ sec.  Assuming reheating
temperatures $\trh \agt m_{B^1} / 25$, the $B^1$'s are produced with
significant thermal relic densities, and these models violate the
bounds of \figref{diffuseflux}.  This region of the phase diagram is
therefore also excluded given standard cosmological assumptions.
However, as in \secref{charged}, for lower reheat temperatures, bounds
from the diffuse photon flux may be evaded.

Our bound assumes that the $B^1$ decays during matter domination, and
we have neglected the impact of the clustering around dark matter
halos. Figure~\ref{fig:diffuseflux} shows that the larger the $B^1$
lifetime $\tau$ the more stringent the constraint on $\Omega_{B^1}$.
Of course, for extremely long lifetimes $\tau$ greater than the age of
the Universe, the constraint weakens again.  For $\tau \agt t_0$, it
is possible to derive a strict bound from diffuse particles in the
galactic halo, which would also contribute to the diffuse gamma ray
background~\cite{inprogress}.

\subsection{WIMP Thermal Relic Density}
\label{sec:thermal}

In Phases 1 and 2 with $R^{-1} > 810~\gev$, the $B^1$ is lighter than
the $G^1$, and so it is absolutely stable and a WIMP dark matter
candidate.  Given standard cosmological assumptions, the $B^1$ thermal
relic may then be determined. The first calculation of
Ref.~\cite{Servant:2002aq} has now been refined by the inclusion of
the full mUED spectrum, radiative corrections to KK particle masses,
$n=2$ resonances, and all co-annihilation
processes~\cite{Kakizaki:2005en,Kakizaki:2005uy,Burnell:2005hm,%
Kong:2005hn,Kakizaki:2006dz}.

The results of Ref.~\cite{Kakizaki:2006dz} are reproduced in
\figref{summary}, which is discussed more fully in \secref{summary}.
Requiring that the $B^1$ thermal relic density not exceed the observed
dark matter density provides yet another cosmological constraint on
the mUED phase diagram, excluding the lower right-hand portion of the
$(R^{-1}, m_h)$ plane.

Of course, as with the other cosmological constraints, the WIMP relic
density constraint may also be avoided by assuming a lower reheating
temperature, which, in this case, is somewhat below $m_{B^1} / 25$.

\section{Summary and LHC Signals}
\label{sec:summary}

In this study we have analyzed the complete parameter space of minimal
universal extra dimensions (mUED), in many senses the simplest extra
dimensional extension of the standard model.  In mUED all particles
propagate in one additional flat, compact dimension, and the model
introduces only 1 additional free parameter, the compactification
radius $R$ of the extra dimension.  Despite this extremely simple
structure, we find that mUED encompasses a wide variety of seemingly
exotic and spectacular predictions for particle colliders and
cosmology, once the entire parameter space is considered.

\begin{figure}
\resizebox{4.56in}{!}{
\includegraphics{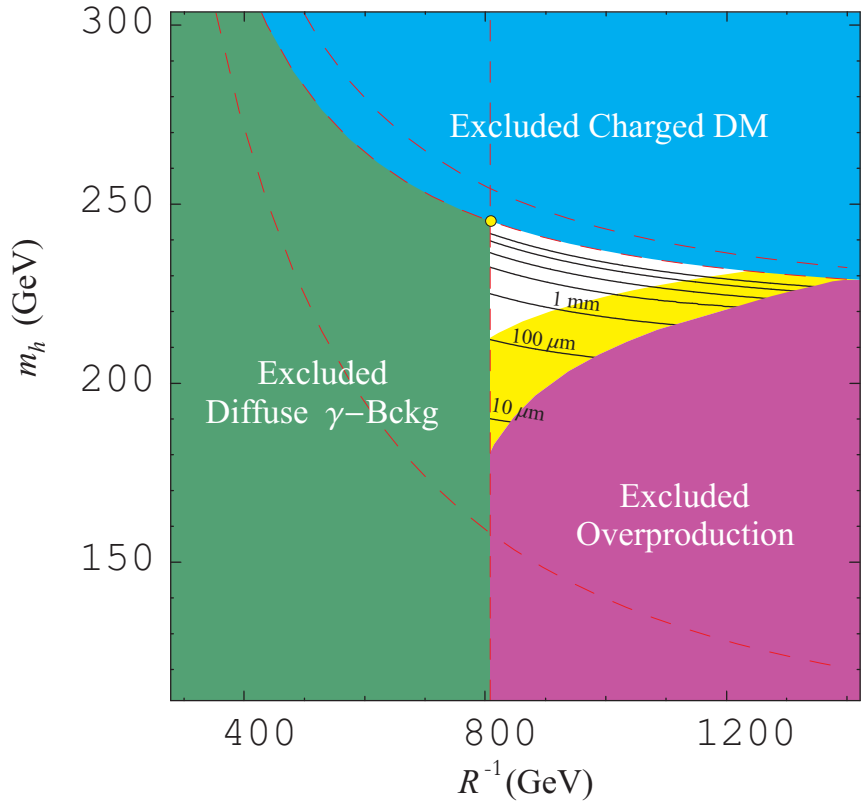}
} 
\caption{The cosmologically preferred region of the complete phase
diagram of mUED.  The $G^1$ has been included, and the dark shaded
regions are excluded by the cosmological constraints on stable charged
relics, the diffuse photon flux, and WIMP overproduction, as
indicated.  In the preferred region, the light shaded region is from
Ref.~\cite{Kakizaki:2006dz} and shows where the $B^1$ thermal relic
density is within $2\sigma$ of the WMAP central value for non-baryonic
dark matter.  Contours of constant decay length $c \tau = 10~\mu\m,
100~\mu\m, 1~\text{mm}, \ldots, 1~\m, 10~\m$ are also plotted (only
the lowest few are labeled).
\label{fig:summary} }
\end{figure}

Our results are summarized in Figs.~\ref{fig:phases},
\ref{fig:lifetimes}, and \ref{fig:summary}.  We began by setting aside
the KK graviton $G^1$ and cosmological considerations, leading to the
complete collider phase diagram of \figref{phases}.  We find that mUED
supports 4 distinct standard model (NLKP, LKP) combinations, or
phases, with qualitatively different implications for signatures.
Potentially most spectacular is the prediction of long-lived particles
at colliders.  The NLKP decay lengths throughout parameter space are
given in \figref{lifetimes}. Long-lived TeV-scale particles might
appear exotic and unlikely.  However, the example of mUED provides a
simple, concrete counterexample that highlights a generic possibility:
in any theory where TeV-scale particles receive identical tree-level
mass contributions $M$ from some new source, the typical splittings
one might expect from radiative or electroweak symmetry breaking
effects are of the order of $M / (16 \pi^2), m_W^2/M \sim 10~\gev$.
Modest additional cancellations can bring this down to $\sim 1~\gev$,
and such splittings lead to macroscopic decay lengths in three-body
decays.

In mUED, the KK spectrum is highly degenerate, and so
strongly-interacting KK particles will be produced with large rates at
the LHC.  Long-lived NLKP tracks will therefore presumably be most
easily identified in the cascades decays of KK quarks and gluons.
Such events will be characterized by many jets and missing transverse
energy, which will satisfy trigger criteria, and the jets will fix the
interaction point.  The possible signals are:
\begin{itemize}
\item Phase 1: Prompt decays $l_R^1 \to B^1 l_R$, where $l = e, \mu,
  \tau$, the mass splitting between KK states is $\Delta m \sim {\cal
  O}(\gev)$, and the final state lepton is consequently very soft.
\item Phase 2: Decays $\kkh \to B^1 f \bar{f}'$, where $f \bar{f}' =
  e^+ \nu_e, \mu^+ \nu_{\mu}, u \bar{d}, \tau^+ \nu_{\tau}, c
  \bar{s}$, where the decay length is $c \tau \agt 100~\text{nm}$ (for
  $R^{-1} \alt 1400~\gev$) and may be effectively infinite for
  collider phenomenology.  Again $\Delta m \sim {\cal O}(\gev)$, and
  the final state fermions are very soft.  Depending on the
  observability of the final state fermions, the exotic signatures
  could include non-prompt decays producing displaced vertices, tracks
  with non-vanishing impact parameters, track kinks, or even
  disappearing charged ($\kkh$) tracks that mysteriously vanish after
  passing through only part of the detector.  In the parameter region
  where the $\kkh$ is effectively stable, it may be produced at low
  velocities, resulting in time-of-flight anomalies and
  highly-ionizing tracks.
\item Phase 3: Decays $B^1 \to \kkh f \bar{f}'$, where the $f
  \bar{f}'$ pairs are as in Phase 2, with decay length typically
  satisfying $c \tau \agt 10~\text{mm}$ (except in a tiny region, in
  which could be even shorter than $10 \mu\m$), and again $\Delta m
  \sim {\cal O}(\gev)$, and the final state standard model fermions
  are very soft. The possible signatures are as above, with the
  exception that, since the NLKP is neutral and the LKP is charged in
  this case, NLKP events could instead be seen as charged ($\kkh$)
  tracks that mysteriously {\em appear} somewhere in the detector.
\item Phase 4: Decays $A^1 \to \kkh f \bar{f}$, where $f = e, \mu,
  \tau, u, d, s, c$, and the decay length is constrained to the
  relatively narrow range $10~\mu\m \alt c \tau \alt 1~\text{mm}$.
  The signatures are as in Phase 3.
\end{itemize}

Following these collider results, we re-introduced the KK graviton
$G^1$ and considered cosmological constraints.  For a low enough
reheating temperature after inflation, (portions of) all 4 phases,
even Phases 3 and 4 with charged LKPs, were viable, justifying our
efforts to classify their collider signatures.

At the same time, much of the phase diagram is excluded if one assumes
a standard cosmology with reheat temperature above $R^{-1}/25$.  The
final results are given in \figref{summary}. Phases 3 and 4 are
excluded by bounds on stable charged particles, Phases 1 and 2 with
$R^{-1} < 810~\gev$ are excluded by bounds from the observed diffuse
MeV photon flux, and Phases 1 and 2 with high $R^{-1}$ are excluded
because WIMPs are overproduced through thermal freeze-out.  

The resulting cosmologically preferred region is bounded on all sides,
resulting in the ``triangle'' shown in \figref{summary}.  In this
region
\begin{itemize}
\item The Higgs boson mass lies in the range $180~\gev \alt m_h \alt
  245~\gev$.  This region is allowed by indirect bounds on $m_h$, and
  implies the ``golden'' 4 lepton signatures for Higgs bosons at the
  LHC.
\item The compactification radius satisfies $810~\gev \alt R^{-1} \alt
  1400~\gev$.  The LKP mass is therefore in this range, and the
  splitting between the LKP and the heaviest $n=1$ KK particle, the KK
  gluon $g^1$, is never more than 320 GeV.  KK particles will
  therefore be copiously produced at the LHC. On the contrary, none of
  these new particles would be produced directly at the International
  Linear Collider operating at center of mass energies below 1.5 TeV.
\item The NLKP to LKP decay is $\kkh \to B^1 f \bar{f}'$, with decay
  lengths satisfying $c\tau \agt 4~\mu\m$, with effectively no upper
  bound.  Generically, then, long-lived tracks are predicted for the
  LHC, leading to the wealth of novel signatures detailed above.
\end{itemize}
These features differentiate mUED from essentially all other proposals
for new electroweak physics.  In particular, they differ markedly from
supersymmetry, in which all of these features would be viewed as
extraordinarily unnatural.
  
We have not examined the cross sections for, and backgrounds to, NLKP
production at the LHC in these various phases.  Such an analysis is,
of course, required if one is to conclude anything about the
observability of these interesting phenomena.  We have also not
considered the extension of mUED to include neutrino masses and KK
right-handed neutrinos~\cite{Matsumoto:2006bf}.  The existence of
these new states at mass $R^{-1}$ provides an alternative
non-gravitational decay to the decays to $G^1$ discussed here.  Given
the simplicity of mUED and the results obtained here, all of these
directions merit further investigation.

We note also that the mUED scenario provides a simple particle physics
framework for seemingly exotic cosmology. For example, it is
remarkable that the values of $R^{-1}$ that give significant $B^1$
thermal relic densities also give extremely degenerate $B^1$ and $G^1$
states. Throughout the mUED parameter space, it is quite easy to
envision scenarios in which the dark matter is produced from the
decays of cold, thermal relics with a wide variety of lifetimes. For
example, we can consider decays at $\sim 10^6~\s$, as in the superWIMP
framework, or during the matter-dominated epoch of the universe, as in
metaCDM ~\cite{Strigari:2006jf}, or even lifetimes longer than the age
of the universe. In the context of cosmological small-scale structure,
dark matter from decays may help alleviate problems facing standard
CDM models~\cite{Sigurdson:2003vy,Cembranos:2005us,Kaplinghat:2005sy,%
Strigari:2006jf}.  These unique phenomenological signatures suggest
new avenues for investigating the identity of dark matter.

\begin{acknowledgments}
We thank H.-C.~Cheng, K.~Matchev, A.~Rajaraman and F.~Takayama for
stimulating correspondence and conversations.  The work of JARC and
JLF is supported in part by NSF CAREER grant No.~PHY--0239817, NASA
Grant No.~NNG05GG44G, and the Alfred P.~Sloan Foundation. The work of
JARC is also supported by the FPA 2005-02327 project (DGICYT, Spain).
LES is supported in part by a Gary McCue Postdoctoral Fellowship
through the Center for Cosmology at UC Irvine.
\end{acknowledgments}

\appendix

\section{Decay Width Analysis}
\label{sec:appendix}

\subsection{Notation and Approximations}

We use the following notation for the electromagnetic fine structure
constant, the weak mixing angle, the hypercharge and weak coupling
constants, and the 4-dimensional Planck mass:
\begin{eqnarray}
\alpha = \frac{e^2}{4\pi} &\simeq& \frac{1}{137} \ , \ \frac{1}{128} \\
\sin^2 \theta_W &\simeq& 0.238 \ , \ 0.231 \\
g' &=& \frac{e}{\cos \theta_W} \\
g &=& \frac{e}{\sin \theta_W} \\
M_4 &=& \frac{1}{\sqrt{16 \pi G_N}} = \frac{1}{\sqrt{2}} M_* 
\simeq 1.72 \times 10^{18}~\gev \ .
\end{eqnarray}
In the first 2 lines, the first numerical values given are appropriate
for processes with momentum transfer of $\sim 1~\gev$, and the second
values are those at the weak scale appropriate for evaluation of the
KK mass spectrum.

Throughout this paper, we consider decays $X(M, p) \to Y(m_1, q_1)
Z_2(m_2, q_2) [Z_3(m_3, q_3)]$, with mass and momentum labels given,
where $X$ and $Y$ are heavy KK level $n=1$ particles and $Z_2$ and
$Z_3$ are light KK level $n=0$ (standard model) particles.  We neglect
the effects of standard model particle masses in all open decay
channels.  With this approximation, the two- and three-body decay
widths are
\begin{eqnarray}
\Gamma_2 &=& \frac{1}{16\pi} \frac{M^2 - m_1^2}{M^3} \,
\overline{| {\cal M} |^2} \\
\Gamma_3 &=& \int_{m_1^2}^{M^2} dm_{12}^2 
\int_0^{\frac{(M^2 - m_{12}^2)(m_{12}^2 - m_1^2)}{m_{12}^2}}
dm_{23}^2 \, \frac{1}{256 \pi^3 M^3} \,
\overline{| {\cal M} |^2} \ ,
\end{eqnarray}
respectively, where $m_{ij}^2 \equiv (q_i + q_j)^2$. 

In presenting the results for decay widths, it is convenient to define
\begin{eqnarray}
\Delta m &\equiv& M - m_1 \\
y &\equiv& \frac{m_1^2}{M^2} \\
N_C &=& \sum_i N_C^i \ ,
\end{eqnarray}
where $N_C$ is the sum of color factors over all kinematically
accessible channels.  For three-body decays, we include only diagrams
mediated by off-shell standard model gauge bosons, and neglect all
others, which are mediated by much heavier KK particles and are
suppressed by small Yukawa couplings.  For example, for $\kkh \to B^1
f \bar{f}'$, we include the contribution from $\kkh \to B^1 W^{\pm\,
*} \to B^1 f \bar{f}'$ but neglect the contributions from $\kkh \to f
\bar{f}'^{1\, *}, \bar{f}' f^{1\, *} \to B^1 f \bar{f}'$.  For all
three-body decays, we also assume $\Delta m \ll m_W$.

Finally, in the width formulae below, the symbol $\approx$ appears
before expressions that are valid assuming $\Delta m \ll M$.

\subsection{Two-body Non-gravitational Decays}

Feynman rules for the relevant vertices are
\begin{eqnarray}
l_R^n B^m_{\mu} \bar{l}_R^{n-m} : && -i g' \gamma_{\mu} \\
H^{- \, n} \bar{l}_R^m \nu_l^{n-m} : 
&& -i \frac{g}{\sqrt{2}} \, \frac{m_l}{m_W} P_L \\
A^n \bar{l}_R^m l_L^{n-m} : 
&& -i \frac{g}{\sqrt{2}} \, \frac{m_l}{m_W} \gamma^5 P_L \ .
\end{eqnarray}

The decay widths are
\begin{eqnarray}
\Gamma(l_R^1 \to B^1 l_R) &=&
\frac{g'^2}{16 \pi} \frac{M}{y} (1-y)^2 (1 + 2y) \\ 
&\approx& 
\frac{3 \alpha}{\cos^2 \theta_W} \frac{ (\Delta m)^2}{M} \\
&\simeq& 2.87 \times 10^{-5}~\gev 
\left[ \frac{\Delta m}{\gev} \right]^2
\left[ \frac{\tev}{M} \right] \\
&\simeq& \left[ 6.87 \times 10^{-12}~\m
\left[ \frac{\gev}{\Delta m} \right]^2
\left[ \frac{M}{\tev} \right] \ \right]^{-1}
\label{decaylb}
\end{eqnarray}
and
\begin{eqnarray}
\Gamma(H^{+ \, 1} \to \bar{l}_R^1 \nu_l) 
&=& \Gamma(A^1 \to \bar{l}_R^1 l_L) \\
&=& \frac{g^2}{32\pi} \frac{m_l^2}{m_W^2} M (1-y)^2 \\ 
&\approx& 
\frac{\alpha}{2 \sin^2 \theta_W} \frac{m_l^2}{m_W^2} 
\frac{(\Delta m)^2}{M} \\
&\simeq& 8.10 \times 10^{-9}~\gev \, \frac{m_l^2}{m_{\tau}^2}
\left[ \frac{\Delta m}{\gev} \right]^2
\left[ \frac{\tev}{M} \right] \\
&\simeq& \left[ 2.44 \times 10^{-8}~\m \, \frac{m_{\tau}^2}{m_l^2}
\left[ \frac{\gev}{\Delta m} \right]^2
\left[ \frac{M}{\tev} \right] \ \right]^{-1} \ .
\label{decayhl}
\end{eqnarray}

\subsection{Three-body Non-gravitational Decays}

Feynman rules for the relevant KK vertices are
\begin{eqnarray}
H^{\pm\, n} B^n W^{\mp} : && i \frac{g'}{2} m_W \\
A^n(p) H^{\pm\, n}(q_1) W_{\mu}^{\mp} : && \frac{g}{2} (p+q_1)_{\mu} \ .
\end{eqnarray}

The decay widths are
\begin{eqnarray}
\Gamma(\kkh \to B^1 f \bar{f}') &=& 
\frac{N_C g^2 g'^2}{49152 \pi^3} \, \frac{M^5}{m_W^2 m_1^2}
\times \nonumber \\
&& \left[ (1-y) (1 + y + 73y^2 + 9y^3) 
 + 12 y^2 (3 + 4y) \ln y \right] \\ 
&\approx& 
\frac{ N_C \alpha^2}{80 \pi \sin^2 \theta_W \cos^2 \theta_W} \,
\frac{(\Delta m)^5}{m_W^2 M^2} \\
&\simeq& 1.96 \times 10^{-16}~\gev \, N_C
\left[ \frac{\Delta m}{\gev} \right]^5
\left[ \frac{\tev}{M} \right]^2 \\
&\simeq& \left[ 1.01~\m \, \frac{1}{N_C}
\left[ \frac{\gev}{\Delta m} \right]^5
\left[ \frac{M}{\tev} \right]^2 \ \right]^{-1} \ , 
\label{decayhb} \\
\Gamma(B^1 \to \kkh f \bar{f}') &=& 
\frac{N_C g^2 g'^2}{147456 \pi^3} \, \frac{M^3}{m_W^2}
\times \nonumber \\
&& \left[ (1-y) (9 + 73y + y^2 + y^3) 
 + 12 y^2 (4 + 3y) \ln y \right] \\ 
&\approx& 
\frac{ N_C \alpha^2}{240 \pi \sin^2 \theta_W \cos^2 \theta_W} \,
\frac{(\Delta m)^5}{m_W^2 M^2} \\
&\simeq& 6.52 \times 10^{-17}~\gev \, N_C
\left[ \frac{\Delta m}{\gev} \right]^5
\left[ \frac{\tev}{M} \right]^2 \\
&\simeq& \left[ 3.03~\m \, \frac{1}{N_C}
\left[ \frac{\gev}{\Delta m} \right]^5
\left[ \frac{M}{\tev} \right]^2 \ \right]^{-1} \ ,
\label{decaybh} 
\end{eqnarray}
and
\begin{eqnarray}
\Gamma(A^1 \to \kkh f \bar{f}') &=& 
\frac{N_C g^4}{12288 \pi^3} \, \frac{M^5}{m_W^4}
\left[ (1-y) (1 - 7y - 7y^2 + y^3) - 12 y^2 \ln y \right] \\ 
&\approx& 
\frac{ N_C \alpha^2}{60 \pi \sin^4 \theta_W} \,
\frac{(\Delta m)^5}{m_W^4} \\
&\simeq& 1.40 \times 10^{-13}~\gev \, N_C
\left[ \frac{\Delta m}{\gev} \right]^5 \\
&\simeq& \left[ 1.41~\mm \, \frac{1}{N_C}
\left[ \frac{\gev}{\Delta m} \right]^5 \ \right]^{-1} \ .
\label{decayah}
\end{eqnarray}

\subsection{Gravitational Interactions}

Feynman rules for gravitons in UED have been presented in
Ref.~\cite{Feng:2003nr}.  Here we abstract those most relevant for the
phenomenology of the lightest KK states.

The graviton interactions are given by 
\begin{equation}
{\cal L}_{\text{int}} = \sum_n \frac{1}{2 M_4} 
G^n_{\mu\nu} T^{n\, \mu\nu}_+ \ ,
\end{equation}
where the sum is over KK levels.  The stress-energy tensor receives
contributions from scalars, fermions, and gauge bosons of the form
\begin{eqnarray}
T^{n\, \mu\nu}_{+\, H} &=& \sum_{m=0}^n 
\left[ \left( \eta^{\mu \rho} \eta^{\nu\sigma} +
\eta^{\mu\sigma} \eta^{\nu\rho} - \eta^{\mu\nu} \eta^{\sigma\rho}
\right) D_{\rho} H^{m \, \dagger} D_{\sigma} H^{n-m} \right. \nonumber \\
&& \qquad + \left. \eta^{\mu\nu} m_H^2 H^{m\, \dagger} H^{n-m} \right] \\
T^{n\, \mu\nu}_{+\, \psi} &=& \sum_{m=0}^n \biggl[
\eta^{\mu\nu}(\overline{\psi_L^m}
i\gamma^{\rho}D_{\rho}\psi_L^{n-m} -m_{n-m}
\overline{\psi_R^m}\psi_L^{n-m}) \nonumber \\
&&\qquad 
- \frac{1}{2} \overline{\psi^m_L} i \gamma^{\mu} D^{\nu} \psi^{n-m}_L
- \frac{1}{2} \overline{\psi^m_L} i \gamma^{\nu} D^{\mu} \psi_L^{n-m}
\nonumber\\
&&\qquad - \frac{1}{2}\eta^{\mu\nu}\partial^{\rho}
(\overline{\psi^m_L}i\gamma_{\rho}\psi_L^{n-m})
+\frac{1}{2}\eta^{\mu\nu}(m_m+m_{n-m})
(\overline{\psi^m_R}\psi_L^{n-m})\nonumber\\
&&\qquad + \frac{1}{4}\partial^{\mu}
(\overline{\psi_L^m}i\gamma^{\nu}\psi_L^{n-m})
+\frac{1}{4}\partial^{\nu}
(\overline{\psi_L^m}i\gamma^{\mu}\psi_L^{n-m})+(R\leftrightarrow L)
\biggr] \\
T^{n\, \mu\nu}_{+\, B} &=& \sum_{m=0}^n\left[ F^{m\, \mu\rho}
F^{n-m\, \nu}{}_{\rho} - \frac{1}{4} \eta^{\mu\nu} F^m_{\rho\sigma}
F^{n-m\, \rho\sigma} \right. \nonumber \\
&& \qquad \left. + m_n m_{n-m} (B^{m\, \mu} B^{n-m\, \nu}
-\frac{1}{2}\eta^{\mu\nu}B_{\rho}^n B^{n-m\, \rho})\right] ,
\end{eqnarray}
where $\psi^0_R(x)=0$, $m_n = n/R$, $D$ is the covariant derivative,
and $F^m_{\mu\nu} \equiv \partial_{\mu} B^m_{\nu} - \partial_{\nu}
B^m_{\mu}$.

The sum over graviton polarizations is
\begin{eqnarray}
\sum_i \epsilon_{\mu\nu}^{n\, i} (k) \epsilon_{\sigma\rho}^{n\, i *} (k)
&=& 
2\left(\eta_{\mu\rho}-\frac{k_\mu
k_\rho}{m_n^2}\right) \left(\eta_{\nu\sigma}-\frac{k_\nu
k_\sigma}{m_n^2}\right) +2\left(\eta_{\mu\sigma}-\frac{k_\mu
k_\sigma}{m_n^2}\right)
\left(\eta_{\nu\rho}-{k_\nu k_\rho\over m_n^2}\right)\nonumber\\
&& - \frac{4}{3} \left(\eta_{\mu\nu}-\frac{k_\mu
k_\nu}{m_n^2}\right) \left(\eta_{\rho\sigma}-\frac{k_\rho
k_\sigma}{m_n^2}\right)\ .
\label{B}
\end{eqnarray}

\subsection{Two-body gravitational Decays}

Feynman rules for the relevant KK vertices are
\begin{eqnarray}
G^n_{\mu\nu} \bar{\psi}^m(p) \psi^{n-m}(q_2) : &&
i \frac{1}{4M_4}\biggl[ \eta_{\mu\nu}
\left[ (\gamma^{\rho}q_{2\, \rho}-m_{n-m})-(\gamma^{\rho}p_{\rho}-m_m) 
\right] \nonumber\\
&&  \quad -\frac{1}{2}(p+q_2)_{\mu}\gamma_{\nu}-
\frac{1}{2}(p+q_2)_{\nu}\gamma_{\mu}\biggr] \\
G^n_{\mu\nu} B^m_{\alpha}(p)
B^{n-m}_{\beta}(q_2) : && i \frac{1}{2M_4}\biggl[\eta_{\alpha\beta}
p_{\mu} q_{2\, \nu}
-\eta_{\mu\alpha} p_{\beta}q_{2\, \nu} 
- \eta_{\nu\beta} p_{\mu}q_{2\, \alpha}
+\eta_{\mu\alpha} \eta_{\nu\beta} (p \cdot q_2) \nonumber \\
&& -\frac{1}{2}\eta_{\mu\nu} \left( \eta_{\alpha\beta}(p \cdot q_2)
 -p_{\beta}q_{2\, \alpha} \right) +
m_n m_{n-m} ( \eta_{\mu\alpha}\eta_{\nu\beta}
-\frac{1}{2} \eta_{\mu\nu}\eta_{\alpha\beta} ) \nonumber \\
&& + \left(\alpha \leftrightarrow \beta \right) \biggr] \ .
\end{eqnarray}

The decay widths are
\begin{eqnarray}
\Gamma(l_R^1 \to G^1 l_R) &=&
\frac{1}{96\pi} \frac{M^7}{M_4^2 m_1^4} (1-y)^4 (2 + 3 y) \\ 
&\approx& 
\frac{5}{6\pi} \frac{ (\Delta m)^4}{M_4^2 M} \\
&\simeq& 8.97 \times 10^{-41}~\gev 
\left[ \frac{\Delta m}{\gev} \right]^4
\left[ \frac{\tev}{M} \right] \\
&\simeq& \left[ 7.34 \times 10^{15}~\s
\left[ \frac{\gev}{\Delta m} \right]^4
\left[ \frac{M}{\tev} \right] \ \right]^{-1}
\label{decaylg}
\end{eqnarray}
and
\begin{eqnarray}
\Gamma(B^1 \to G^1 \gamma) &=&
\frac{\cos^2 \theta_W}{144\pi} \frac{M^7}{M_4^2 m_1^4} 
(1-y)^3 (1 + 3 y + 6 y^2) \\ 
&\approx& 
\frac{5 \cos^2 \theta_W}{9\pi} \frac{ (\Delta m)^3}{M_4^2} \\
&\simeq& 4.55 \times 10^{-38}~\gev 
\left[ \frac{\Delta m}{\gev} \right]^3 \\
&\simeq& \left[ 1.45 \times 10^{13}~\s
\left[ \frac{\gev}{\Delta m} \right]^3 \ \right]^{-1} \ .
\label{decaybg}
\end{eqnarray}

\subsection{Three-body Gravitational Decays}

The relevant KK interaction vertex is
\begin{equation}
H^{\pm\, n}(p) G^n_{\mu\nu} W_{\rho}^{\mp} : i \frac{m_W}{2M_4} 
(\eta_{\mu\rho} \eta_{\nu\sigma} + \eta_{\mu\sigma} \eta_{\nu\rho}
- \eta_{\mu\nu} \eta_{\rho\sigma} ) p^{\sigma} \ .
\end{equation}

The decay width is
\begin{eqnarray}
\Gamma(\kkh \to G^1 f \bar{f}') &=& 
\frac{N_C \, g^2}{13824 \pi^3} \, \frac{M^9}{m_W^2 M_4^2 m_1^4}
\left[ (1-y) (1 - 5y - 5y^2 \right. \nonumber \\
&& \left. - 245y^3 - 50y^4 + 4y^5) 
 - 60 y^3 (2 + 3y) \ln y \right] \\
&\approx&
\frac{N_C \ 2 \alpha}{63 \pi^2 \sin^2 \theta_W} \,
\frac{(\Delta m)^7}{m_W^2 M_4^2 M^2} \\
&\simeq& 5.58 \times 10^{-51}~\gev \,  
{N_C} \left[ \frac{\Delta m}{\gev} \right]^7
\left[ \frac{\tev}{M} \right]^2 \\
&\simeq& \left[ 1.18 \times 10^{26}~\s \,  
\frac{1}{N_C} \left[ \frac{\gev}{\Delta m} \right]^7
\left[ \frac{M}{\tev} \right]^2 \ \right]^{-1} \  .
\label{decayhg}
\end{eqnarray}

%%%%%%%%%%%%%%%%%%%%%%%%%%%%%%%%%%%%%%%%%%%%%%%%%%%%%%

\end{document}